\documentclass[a4paper,10pt,prl,twocolumn,showpacs,aps,floats,superscriptaddress]{revtex4}

\usepackage{epsfig}

\begin{document}
\noindent [Phys. Rev. E {\bf 68}, 035103(R) (2003)]

\title{Epidemic Incidence in Correlated Complex Networks} 

\author{Yamir Moreno}

\affiliation{Departamento de F\'{\i}sica Te\'orica, Universidad de
Zaragoza, Zaragoza 50009, Spain}

\affiliation{Instituto de Biocomputaci\'on y F\'{\i}sica de Sistemas
Complejos, Universidad de Zaragoza, Zaragoza 50009, Spain}

\author{Javier B. G\'omez} 

\affiliation{Departamento de Ciencias de la Tierra, Universidad de
Zaragoza, Zaragoza 50009, Spain}

\author{Amalio F. Pacheco} 

\affiliation{Departamento de F\'{\i}sica Te\'orica, Universidad de
Zaragoza, Zaragoza 50009, Spain}

\affiliation{Instituto de Biocomputaci\'on y F\'{\i}sica de Sistemas
Complejos, Universidad de Zaragoza, Zaragoza 50009, Spain}

\date{\today}

\widetext

\begin{abstract} 

We introduce a numerical method to solve epidemic models on the
underlying topology of complex networks. The approach exploits the
mean-field like rate equations describing the system and allows to work
with very large system sizes, where Monte Carlo simulations are
useless due to memory needs. We then study the SIR epidemiological
model on assortative networks, providing numerical evidence of the
absence of epidemic thresholds. Besides, the time profiles of the
populations are analyzed. Finally, we stress that the present method
would allow to solve arbitrary epidemic-like models provided that they
can be described by mean-field rate equations.

\end{abstract}

\pacs{89.75.-k, 89.75.Fb, 05.70.Jk, 05.40.a}

\maketitle


A few years ago, Watts and Strogatz \cite{ws98} introduced a model
able to produce networks with properties of both regular lattices and
random graphs with small diameter. Their model soon led to a burst of
activity in the field \cite{strogatz,doro}, further spurred by Barabasi and
collaborators who found that many seemingly diverse systems share several
topological properties such as a power law behavior in their
connectivity distributions when represented as networks
\cite{bar99}. These complex networks are formed by a set of many
elements (or nodes) that are linked together through edges (or links)
if they interact directly. Empirical evidence supports that in notable
networks, such as metabolic or communication webs, the probability
$P(k)$ that any node has $k$ links to other nodes is distributed
accordingly to a power law $P(k)\sim k^{-\gamma}$
\cite{bara02,fal,romu01}, with $\gamma \le 3$ in most cases.
 
Networks of this type, called scale-free (SF) networks, show a
noticeable property: the heterogeneity of the connectivity
distribution can not be neglected. One of the fundamental results
derived from this property is that the threshold characterizing the
percolation transition or an epidemic outbreak is vanishing in the
thermodynamic limit \cite{havlin01,newman00,bar00,pv01a}. On the other
hand, many real-world networks are also characterized by degree
correlations \cite{n02a,kr01}, which make it necessary to reconsider
the same problems but taking into account the conditional probability
$P(k'|k)$ that a link emanating from a node of connectivity (or
degree) $k$ leads to a node of connectivity $k'$. Very recently, it
has been shown analytically that the presence of nontrivial
correlations does not change the main conclusions drawn for
uncorrelated graphs, namely, the absence of percolation and epidemic
thresholds \cite{av03,brv03} under very general conditions. However,
neither complete exact solutions nor numerical studies of the system's
dynamics have been reported. From a theoretical side, it seems an
unsurmountable task to work out such study. In contrast, although
feasible in principle, numerical simulations using Monte Carlo
techniques would be quite arduous since one should first generate a
network with the proper correlations and then perform expensive Monte
Carlo simulations for large system sizes, where memory requirements
could seriously limit the size of the networks under study.

In this paper, we present an efficient numerical method that allows
the study of epidemic-like models on the underlying topology of
complex networks with arbitrary connectivity distributions and
degree-degree correlations. The method solves the mean-field rate
equations describing the dynamics of the system, where the topological
properties are accounted for. Specifically, we fully analyze the
Susceptible-Infected-Removed (SIR) model in networks with assortative
correlations, where nodes tend to be linked with their connectivity
peers \cite{n03} as for the case of social networks. Our results are
compared with those obtained for uncorrelated networks and confirm the
previous findings about the absence of any epidemic
threshold. Besides, we report on the time evolution of the individual
populations which allow us to draw interesting conclusions when
confronted to uncorrelated networks.

The SIR model \cite{moreno02,n02b} considers that individuals are
classified in three classes according to their state: susceptible,
infected and removed. The epidemic is propagated by contacts between
infected and susceptible individuals at a rate $\lambda$. Thus, once
an individual gets infected and recovers he can not catch the epidemic
again. Moreover, in networks with power-law distributed connectivities
one also has to consider the presence of nodes with different
connectivity $k$ within each category. We consider the time evolution
of the magnitudes $\rho_k(t)$, $s_k(t)$, and $r_k(t)$, which are the
density of infected, susceptible, and removed nodes of connectivity
$k$ at time $t$, respectively, with the normalization condition
$\rho_k(t)+s_k(t)+r_k(t) = 1$. Global quantities can be obtained by
averaging over the connectivity classes. In this way, the fraction of
infected individuals (or the epidemic incidence) is given by
$r(t)=\sum_k P(k)r_k(t)$.

The mean-field rate equations for the evolution of these densities
satisfy the following set of coupled differential equations \cite{moreno02}:
\begin{eqnarray}
\frac{d s_k(t)}{d t} & = & - \lambda k s_k(t)
\sum_{k'}P(k'|k)\rho_{k'} , \label{eq1}\\
\frac{d \rho_k(t)}{d t} & = & -\rho_k(t) +\lambda k s_k(t)
\sum_{k'}P(k'|k)\rho_{k'}, \label{eq2}\\
\frac{d r_k(t)}{d t} & = & \rho_k(t). \label{eq3}
\end{eqnarray}
with the initial conditions $s(0)=(N-1)/N$, $\rho(0)=1/N$ and
$r(0)=0$. In the above equations, it is considered that infected
individuals recover with unitary rate. Moreover, the creation term in
Eq.\ (\ref{eq2}) is proportional to the density of susceptible
individuals, $s_k(t)$, times the spreading rate $\lambda$, the number
of emanating links $k$ and the probability that any neighboring node
is infected. This probability is given by the average over all degrees
of the probability $P(k'|k)\rho_{k'}$ that a link emanated from a node
with connectivity $k$ points to an infected node with degree $k'$
\cite{brv03,note1}.

The numerical approach introduced here is based on a different
interpretation of the mean-field rate equations\ (\ref{eq1}-\ref{eq3})
and is implemented as follows. Indeed, these equations describe a
process in which individuals are decaying from one state to
another. Hence, one can speak in terms of transition probabilities
from the class of susceptible onto infected and finally onto removed
individuals, {\em i.e.},
$s_k\stackrel{W_{s\rho}^k}{\longrightarrow}\rho_k\stackrel{W_{\rho
r}^k}{\longrightarrow}r_k$. From Eqs.\ (\ref{eq1}-\ref{eq3}) we get
the transition probabilities at each time step $t$:
\begin{eqnarray}
W_{s\rho}^{k}(t) & = & \lambda k N P(k) s_k(t)
\sum_{k'}P(k'|k)\rho_{k'}(t) , \label{eq4}\\
W_{\rho r}^{k}(t) & = & N P(k) \rho_k(t), \label{eq5}
\end{eqnarray}
where all the topological information is contained. The mean time
interval, $\tau$, for one transition to occur after $i-1$ decays is,
\begin{equation}
\tau=\frac{1}{W_{s\rho}(t)+W_{\rho r}(t)}, \label{eq6}
\end{equation}
with $W_{s\rho}(t)=\sum_k W_{s\rho}^{k}(t)$, $W_{\rho r}(t)=\sum_k
W_{\rho r}^{k}(t)$ and $t=\sum_j^{i-1}\tau_j$.  That is, at any
instant $t$ of the decaying process, $\tau$ represents the mean time
for the next individual decay. Once the transition probabilities\
(\ref{eq4}-\ref{eq5}) are calculated, we stochastically decide what
transition takes place. Nodes are divided in three classes according
to their state and, within each of these classes, they are also
characterized by their connectivity $k$. Hence, the identification of
both what transition occurs and which class $k$ is affected after one
$\tau$ is done by deciding that the probabilities that precisely a
node with connectivity $k$ changes its state at time $t$ are given by
\begin{eqnarray}
\Pi_{s\rho}^{k}(t)=W_{s\rho}^{k}(t)\tau, \qquad \qquad
\Pi_{\rho r}^{k}(t)=W_{\rho r}^{k}(t)\tau,  \label{eq7}
\end{eqnarray}
materializing the choice by generating a random number between $0$ and
$1$. In other words, the individual decays within each connectivity
class $k$ proceeds by chance with the probabilities dictated by Eq.\
(\ref{eq7}) until the end of the spreading process,{\em i.e.}, when
the condition $\rho_k=0$ is verified $\forall k$. It is worth
mentioning that the present stochastic approach has been successfully
applied to the study of time-dependent models of fracture
\cite{pre98,jgr}. Besides, we remark that the main advantage of the
method lies in the fact that we don't have to know explicitly the
connectivity matrix, but only a functional form for
$P(k'|k)$. Therefore, in contrast to Monte Carlo simulations, we do
not generate any network. Instead, we generate a sequence of integers
distributed according to a power law $P(k)\sim k^{-\gamma}$ for the
node's connectivities.

\begin{figure}[t]
\begin{center}
\epsfig{file=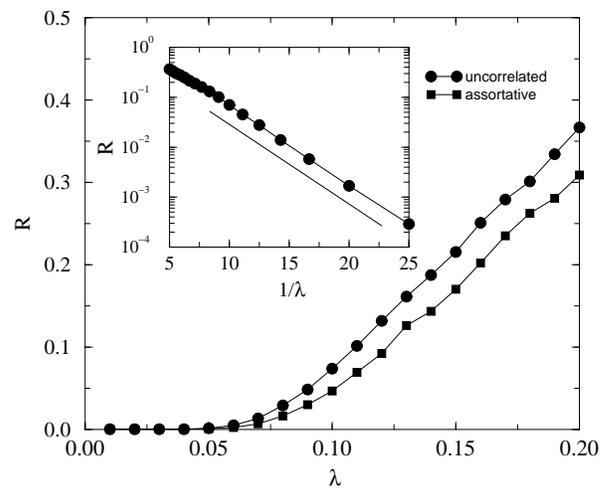,width=2.6in,angle=-90,clip=1}
\end{center}
\caption{Epidemic incidence in scale-free networks with $\gamma=3$ and
  $N=10^5$ with (full squares) and without correlations (full circles)
  as a function of the spreading rate $\lambda$. The assortative
  correlations are given by Eq.\ (\ref{eq9}) with $\alpha=0.1$. The
  inset is a fit to the analytic relation $R\sim e^{-2/ \langle k
  \rangle\lambda}$. See the text for details.}
\label{fig1}
\end{figure}

In order to compare the SIR dynamics in uncorrelated and correlated
networks, we set henceforth $\gamma=3$. For uncorrelated 
networks, the two-point degree correlation function $P(k'|k)$ is of the
form
\begin{equation}
P(k'|k)=q_{k'}=\frac{k'P(k')}{\langle k \rangle}.
\label{eq8}
\end{equation}
Furthermore, consider the case in which the degree
correlations can be decomposed into two components
\begin{equation}
P(k'|k) = (1-\alpha) q_{k'}+ \alpha \delta_{kk'},
\label{eq9}
\end{equation}
with $0\le\alpha<1$. Varying the parameter $\alpha$ one interpolates
between the uncorrelated graphs ($\alpha=0$) and a graph with positive
degree correlations \cite{av03,aw03}. 

\begin{figure}[t]
\begin{center}
\epsfig{file=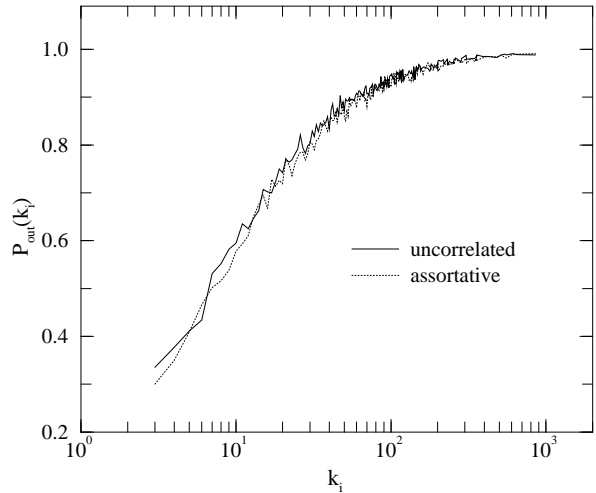,width=2.6in,angle=-90,clip=1}
\end{center}
\caption{Likelihood of an epidemic outbreak as a function of the
  connectivity of the initial seed. The value of $\lambda$ has been
  set to $0.15$. Note that the introduction of correlations does not
  change the probability of having an outbreak. The figure is in
  log-linear scales for clarity. The network's parameters are as of
  Fig.\ \ref{fig1}.}
\label{fig2}
\end{figure}

We have studied the effects introduced by assortative correlations
given by Eq.\ (\ref{eq9}) in the spreading of a disease. Social
networks are the capital example of assortative networks and, although
positively correlated, do not have high correlation coefficients
\cite{n03}, hence, we have used small values of $\alpha$ in our
simulations. Figure\ \ref{fig1} shows the epidemic incidence,
$R=r(t_{\infty})=\sum_k P(k)r_k(t_{\infty})$, averaged over many
realizations, where $t_{\infty}$ is the lifetime of an epidemic
outbreak. The results indicate that the epidemic incidence in networks
with assortative correlations is smaller that for uncorrelated ones
with the same degree distribution. However, the epidemic threshold in
the thermodynamic limit is not modified by the presence of positive
correlations and is vanishing when $N\rightarrow \infty$. The inset in
Fig.\ \ref{fig1} is a plot of $R$ as a function of $1/\lambda$ for an
uncorrelated network with $\langle k \rangle=6$. The behavior nicely
fits the theoretical prediction $R\sim e^{-2/ \langle k
\rangle\lambda}$ \cite{moreno02}, being the slope of the straight line
equal to $0.36(1)$. This constitutes a further test of the validity of
the approach presented here. As to the correlated networks, they
satisfy the same functional form $R\sim e^{-C/\lambda}$ but approaches
zero more slowly. Thus, the numerical findings confirm some
theoretical arguments recently pointed out about the absence of any
epidemic threshold in SF correlated networks with $\gamma\le 3$
\cite{av03,brv03}. Moreover, we note that in finite size networks with
assortative correlations the effective threshold is larger, suggesting
that these networks are more robust than uncorrelated ones.

\begin{figure}[t]
\begin{center}
\epsfig{file=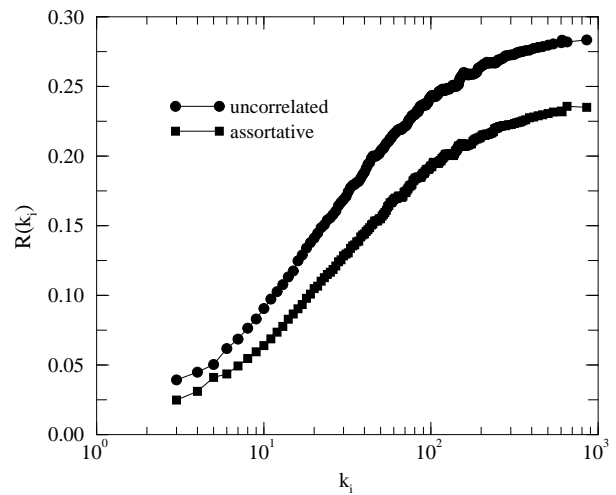,width=2.6in,angle=-90,clip=1}
\end{center}
\caption{Epidemic incidence as a function of the connectivity of the
  initial seed for $\lambda=0.15$. The dependency on the initial seed
  is stronger for uncorrelated networks. The network's parameters are
  as of Fig.\ \ref{fig1}. See the text for further details.}
\label{fig3}
\end{figure}

The high heterogeneity of SF networks also causes that the relative
incidence of an outbreak strongly depends on the connectivity of the
first infected nodes \cite{moreno02}. First, we explored the
likelihood of an epidemic outbreak, $P_{out}$ as a function of the
connectivity, $k_i$, of the initial infected node. This probability is
obtained by dividing the number of times an epidemic developed by the
total attempts made. The results are summarized in Fig.\
\ref{fig2}. As can be observed, no matter whether or not correlations
are present, the probability of having an epidemic outbreak is the
same for both kind of networks. This behavior changes when
coarse-graining the results of Fig.\ \ref{fig1}. The results shown in
Fig.\ \ref{fig3} have been obtained by simulating the SIR dynamics
when the initial infected node has a connectivity $k_i$ and recording
the epidemic incidence in each case. Now, the number of individuals
within the removed class at the end of the spreading process
significantly depends on the correlations. Note that this is not a
consequence of the smaller value of the epidemic incidence for
assortative correlations. In fact, the first part of the curves, up to
intermediate values of $k_i$, follows a logarithmic dependency on the
connectivity, but the number of removed individuals for uncorrelated
networks grows more faster than for correlated ones. This behavior can
be intuitively understood since in assortative networks, nodes are
mainly connected with their connectivity peers (even for small values
of $\alpha$), whereas in uncorrelated networks it is always possible
that a poorly connected node transmits the disease to a hub. In other
words, the random mixing leads to a higher $R$ and favors the
propagation of the outbreak.

\begin{figure}[t]
\begin{center}
\epsfig{file=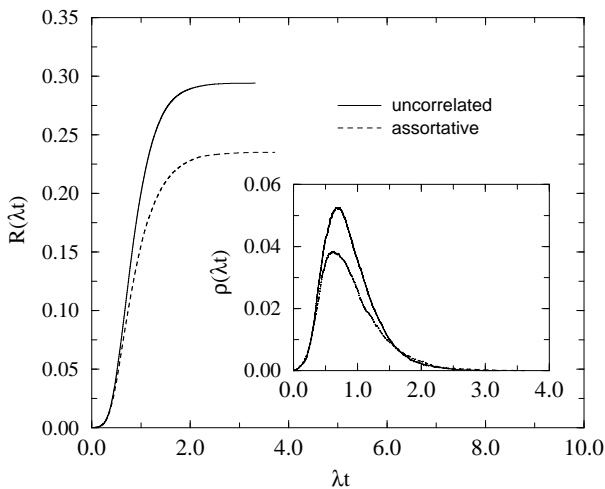,width=2.6in,angle=-90,clip=1}
\end{center}
\caption{Time evolution of the epidemic incidence (main figure) and of
  the density of infected individuals (inset) when the initial
  infected individual has the largest connectivity $k_{max}$ for both
  uncorrelated and correlated networks. $\lambda$ is equal to
  $0.15$. The network's parameters are as of Fig.\ref{fig1}.}
\label{fig4}
\end{figure}

The method introduced here also allows the exploration of the time
dependency of the quantities characterizing the epidemic
spreading. Note that the time profiles, in contrast to MC simulations,
are in units of $\lambda^{-1}$ and thus they can be used as a real
quantitative measure. Figure\ \ref{fig4} plots the time evolution of
the fraction of removed individuals and that of the infected nodes for
the two networks under study when the initial infected individual has
the largest connectivity $k_{max}$. Here, as before, by simple
inspection one would not be able to distinguish the behavior of these
magnitudes in both networks as they show the same functional
dependency on time. However, a more careful look at the plateau of
$R(\lambda t)$ reveals that the epidemic lifetime is longer for
assortative correlations than for uncorrelated networks. This is a
direct consequence of the correlations. Starting from the very hub of
the network, as time goes on, the epidemic is propagated, on average,
from highly connected nodes down to less connected individuals such
that when the end of the process is approaching, the mean time for
individual decays is longer and longer leading to an effective
deceleration of the epidemic spreading process. This is not anymore
the case when random mixing rules out any correlated spreading. A
further evidence of this mechanism is provided in the inset where it
can be clearly noted that the two density profiles crosses well before
the final death of the disease.

In summary, we have introduced an efficient numerical method that
allows to explore the spreading of epidemic diseases in complex
correlated networks without generating explicitly the network with the
proper correlations. We have studied the SIR epidemiological model in
assortative networks and found that its qualitative behavior is the
same as for uncorrelated networks. However, from a more practical
perspective, there are some important quantitative differences that
deserve to be considered more carefully and makes the inclusion of
correlations in nowadays studies meaningful. In particular, while the
likelihood of an epidemic outbreak is not modified when taking into
account positive correlations, the epidemic incidence is smaller than
in networks with no correlations. In large social networks this may
lead to a difference of $15\%$ to $20\%$ of infected people for
moderate values of $\lambda$. On the other hand, we have found that
the diseases are longest-lived in assortative networks.
Additionally, we stress that the method employed here can be used to
solve other epidemic-like models in networks with any correlations
such as the SIS \cite{pv01a} and rumor spreading \cite{ymv03} models,
provided that they can be described through mean-field rate
equations. Finally, we point out that results for
disassortative networks will appear elsewhere.

\begin{acknowledgments}
Y.\ M.\ thanks A. V\'azquez for useful discussions. Y.\
M.\ acknowledges financial support from the Secretar\'{\i}a de Estado
de Educaci\'on y Universidades (Spain, SB2000-0357). This work has
been partially supported by the Spanish DGICYT project BFM2002-01798.
\end{acknowledgments}


\begin{thebibliography}{99}

\bibitem{ws98} D. J. Watts and H. S. Strogatz, Nature {\bf 393}, 440 (1998).

\bibitem{strogatz} S. H. Strogatz, Nature (London) {\bf 410}, 268
(2001).

\bibitem{doro} S. N. Dorogovtsev and J. F. F. Mendes, Adv. Phys. {\bf
51}, 1079 (2002).

\bibitem{bar99} A.-L. Barab\'{a}si, and R. Albert, Science {\bf 286},
509 (1999).

\bibitem{bara02} R. Albert and A.-L. Barab\'{a}si,
Rev. Mod. Phys. {\bf 74}, 47 (2002).

\bibitem{fal} M. Faloutsos, P. Faloutsos, and C. Faloutsos,
Comp. Com. Rev. {\bf 29}, 251 (2000).

\bibitem{romu01} R. Pastor-Satorras, A. V\'{a}zquez, and A. Vespignani,
Phys. Rev. Lett. {\bf 87}, 258701 (2001).

\bibitem{havlin01} R. Cohen, K. Erez, D. ben-Avraham, and
S. Havlin, {\em Phys. Rev. Lett.} {\bf 85}, 4626 (2000);

\bibitem{newman00} D. S. Callaway, M. E. J. Newman, S. H. Strogatz,
and D. J. Watts, {\em Phys. Rev. Lett.} {\bf 85}, 5468 (2000).

\bibitem{bar00} R. Albert, H. Jeong, and A.-L. Barab\'{a}si,
Nature {\bf 406}, 378 (2000).

\bibitem{pv01a} R. Pastor-Satorras, and A. Vespignani,
Phys. Rev. Lett. {\bf 86}, 3200 (2001).

\bibitem{n02a} M. E. J. Newman, Phys. Rev. Lett. {\bf 89}, 208701
	(2002).

\bibitem{kr01} P. L. Krapivsky and S. Redner, Phys. Rev. E {\bf 63},
	066123 (2001).

\bibitem{av03} A. V\'{a}zquez, and Y. Moreno, Phys. Rev. E {\bf 67},
	015101(R) (2003).

\bibitem{brv03} M. Bogu\~n\'a, R. Pastor-Satorras and A. Vespignani,
        Phys. Rev. Lett. {\bf 90}, 028701 (2003).

\bibitem{n03} M. E. J. Newman, Phys. Rev. E {\bf 67}, 026126 (2003). 

\bibitem{moreno02}
	Y. Moreno, R. Pastor-Satorras, and A. Vespignani,
	Eur. Phys. J. B {\bf 26}, 521 (2002).

\bibitem{n02b} 
	M. E. J. Newman, 
	Phys. Rev. E {\bf 66}, 016128 (2002).

\bibitem{note1} We assume that the conditions for the set of equations
  to be meaningful are satisfied \cite{brv03}. 

\bibitem{pre98} J. B. G\'omez, Y. Moreno, and A. F. Pacheco,
Phys. Rev. E {\bf 58}, 1528 (1998).

\bibitem{jgr} Y. Moreno, A. M. Correig, J. B. G\'omez,and
A. F. Pacheco, J. Geophys. Res. {\bf B 106}, 6609 (2001).

\bibitem{aw03} A. V\'{a}zquez, and M. Weigt, Phys. Rev. E {\bf 67},
	027101 (2003).

\bibitem{ymv03} Y. Moreno, M. Nekovee, and A. Vespignani, in
preparation.

\end{thebibliography}
\end{document}